\documentclass[preprint,amsmath,amssymb,superscriptaddress,prb,floatfix,footinbib]{revtex4-1}
\usepackage[]{graphicx}
\usepackage{tabularx}
\usepackage[usenames,dvipsnames]{color}
\usepackage{soul}
\usepackage{bm}

\usepackage{times}
\usepackage{amsfonts}
\usepackage{mathrsfs}
\usepackage{graphicx}
\usepackage{dcolumn}
\usepackage{bm}
\usepackage{color}

\usepackage[colorlinks,bookmarks=false,citecolor=blue,linkcolor=red,urlcolor=blue]{hyperref}

\usepackage{multirow}
\usepackage{physics}
\usepackage{multibib}

\begin{document}






\title{Flat band separation and robust spin-Berry curvature in bilayer kagome metals}

\author{Domenico Di Sante}\email{domenico.disante@unibo.it}
\affiliation{Department of Physics and Astronomy, University of Bologna, 40127 Bologna, Italy}
\affiliation{Center for Computational Quantum Physics, Flatiron Institute, 162 5th Avenue, New York, NY 10010, USA}

\author{Chiara Bigi}
\affiliation{School of Physics and Astronomy, University of St Andrews, St Andrews KY16 9SS, United Kingdom}

\author{Philipp Eck}
\affiliation{Institut f\"{u}r Theoretische Physik und Astrophysik and W\"{u}rzburg-Dresden Cluster of Excellence ct.qmat, Universit\"{a}t W\"{u}rzburg, 97074 W\"{u}rzburg, Germany}

\author{Stefan Enzner}
\affiliation{Institut f\"{u}r Theoretische Physik und Astrophysik and W\"{u}rzburg-Dresden Cluster of Excellence ct.qmat, Universit\"{a}t W\"{u}rzburg, 97074 W\"{u}rzburg, Germany}

\author{Armando Consiglio}
\affiliation{Institut f\"{u}r Theoretische Physik und Astrophysik and W\"{u}rzburg-Dresden Cluster of Excellence ct.qmat, Universit\"{a}t W\"{u}rzburg, 97074 W\"{u}rzburg, Germany}

\author{Ganesh Pokharel}
\affiliation{Materials Department, University of California Santa Barbara, Santa Barbara, California 93106, USA}

\author{Pietro Carrara}
\affiliation{Dipartimento di Fisica, Università degli Studi di Milano, via Celoria 16, 20133 Milano, Italy}
\affiliation{Istituto Officina dei Materiali, Consiglio Nazionale delle Ricerche, Trieste I-34149, Italy}

\author{Pasquale Orgiani}
\affiliation{CNR-IOM TASC Laboratory, Area Science Park, I-34149 Trieste, Italy}

\author{Vincent Polewczyk}
\affiliation{CNR-IOM TASC Laboratory, Area Science Park, I-34149 Trieste, Italy}

\author{Jun Fujii}
\affiliation{Istituto Officina dei Materiali, Consiglio Nazionale delle Ricerche, Trieste I-34149, Italy}

\author{Phil D. C King}
\affiliation{School of Physics and Astronomy, University of St Andrews, St Andrews KY16 9SS, United Kingdom}

\author{Ivana Vobornik}
\affiliation{Istituto Officina dei Materiali, Consiglio Nazionale delle Ricerche, Trieste I-34149, Italy}

\author{Giorgio Rossi}
\affiliation{Dipartimento di Fisica, Università degli Studi di Milano, via Celoria 16, 20133 Milano, Italy}
\affiliation{Istituto Officina dei Materiali, Consiglio Nazionale delle Ricerche, Trieste I-34149, Italy}

\author{Ilija Zeljkovic}
\affiliation{Department of Physics, Boston College, Chestnut Hill, MA 02467, USA}

\author{Stephen D. Wilson}
\affiliation{Materials Department, University of California Santa Barbara, Santa Barbara, California 93106, USA}

\author{Ronny Thomale}
\affiliation{Institut f\"{u}r Theoretische Physik und Astrophysik and W\"{u}rzburg-Dresden Cluster of Excellence ct.qmat, Universit\"{a}t W\"{u}rzburg, 97074 W\"{u}rzburg, Germany}

\author{Giorgio Sangiovanni}\email{sangiovanni@physik.uni-wuerzburg.de}
\affiliation{Institut f\"{u}r Theoretische Physik und Astrophysik and W\"{u}rzburg-Dresden Cluster of Excellence ct.qmat, Universit\"{a}t W\"{u}rzburg, 97074 W\"{u}rzburg, Germany}

\author{Giancarlo Panaccione}\email{panaccione@iom.cnr.it}
\affiliation{Istituto Officina dei Materiali, Consiglio Nazionale delle Ricerche, Trieste I-34149, Italy}

\author{Federico Mazzola}\email{federico.mazzola@unive.it}
\affiliation{Department of Molecular Sciences and Nanosystems, Ca’ Foscari University of Venice, 30172 Venice, Italy}
\affiliation{Istituto Officina dei Materiali, Consiglio Nazionale delle Ricerche, Trieste I-34149, Italy}

\date{\today}

\begin{abstract}
\bf{Kagome materials have emerged as a setting for emergent electronic phenomena that encompass different aspects of symmetry and topology. It is debated whether the XV$_6$Sn$_6$ kagome family (where X is a rare earth element), a recently discovered family of bilayer kagome metals, hosts a topologically non-trivial ground state resulting from the opening of spin-orbit coupling gaps. These states would carry a finite spin-Berry curvature, and topological surface states. Here, we investigate the spin and electronic structure of the XV$_6$Sn$_6$ kagome family. We obtain evidence for a finite spin-Berry curvature contribution at the center of the Brillouin zone, where the nearly flat band detaches from the dispersing Dirac band because of spin-orbit coupling. In addition, the spin-Berry curvature is further investigated in the charge density wave regime of ScV$_6$Sn$_6$, and it is found to be robust against the onset of the temperature-driven ordered phase. Utilizing the sensitivity of angle resolved photoemission spectroscopy to the spin and orbital angular momentum, our work unveils the spin-Berry curvature of topological kagome metals, and helps to define its spectroscopic fingerprint.}
\end{abstract}
\maketitle
Electrons on a kagome lattice constitute a preeminently suited scenario for exotic quantum phenomena at all coupling scales: Within the Mott limit, it is the established paradigmatic setting for spin liquids and other aspects of frustrated magnets~\cite{Balents2010}. For symmetric metallic states or itinerant magnets, the diversity of dispersing kagome signatures such as Dirac cones, flat bands, and van Hove singularities enable to unlock a plethora of correlated electron phenomena from topological band formation and symmetry breaking~\cite{Guo2009,Mazin2014,PhysRevB.86.121105,Kiesel2013,DiSante2020,Ortiz2020,Wu2021,Neupert_2022,Kang_2020, Kang_2020b, Lin_2018, Ye_2019, Sanna_2023}.

The family of XV$_6$Sn$_6$ kagome materials, where X is a rare earth element, belongs to a novel series (hereafter dubbed the “166” family) which has been predicted to host electronic states with non-trivial topology. In particular, not only the surface states that appear at natural cleavage planes of the crystals are theoretically conceived to have a non-trivial origin \cite{Pokharel_2021, Hu_2022}, but also the correlated flat band naturally arising from the kagome geometry~\cite{Mielke_1991} is characterized by a non-zero $\mathbb{Z}_2$ Kane-Mele invariant for the action of spin-orbit coupling (SOC) \cite{Tang_2011, Rhim_2020, Weeks_2012, Gao_2020, Nakai_2022, Yan_2014, Crasto_2019}. If the onsite energy of such a separated flat band could be controlled, one could trigger novel topological phases with potential applications in spintronics and non-volatile electronics \cite{He_2022, Smejkal_2018, He_2019, Alam_2021, Zhang_2015}. Therefore, uncovering the non-trivial topological character of such a flat band would be a true milestone in the field of condensed matter physics.


The direct experimental observation of non-trivial topological properties in 166 kagome metals remains an open challenge. While transport is unable to probe correlated flat band states below E$_F$ or isolate topological surface states \cite{Li_2021, Kang_2020, Chen_2021, Sun_2022, Zhi_2018, Huang_2022, Kang_2020b, Liu_2020, Ye_2018}, in ARPES there are tantalizing hints that these states exist. Crucially, measurements of the spin degree of freedom are missing; When angle-resolved photoemission spectroscopy (ARPES) has been able to detect the surface states manifold and the flat bands in XV$_6$Sn$_6$ systems~\cite{Peng_2021, Hu_2022, Pokharel_2021}, the lack of measurements for the spin degree of freedom and the action on it of time-reversal symmetry hinder the conclusive proof of their topological nature.

Here, we provide the first spectroscopic evidence of the non-trivial topology in the “166” kagome family. We use spin-ARPES and density functional theory (DFT) calculations to determine the electronic structure of these systems resolved in energy, momentum, and spin. We do not only find a net spin polarization of the surface states in the prototypical compound TbV$_6$Sn$_6$, but we ultimately demonstrate the non-trivial topology of the gap between the dispersive Dirac band and the nearly flat band arising from the kagome geometry. Such a gap is a common feature of all kagome lattices with non-zero SOC, and its nature is believed to be topological, yet its demonstration has been elusive until now ~\cite{Liu_2020, Kang_2020, Kang_2020b, Yin_2019}. Remarkably, we detect a finite spin-Berry curvature in kagome metals, and by systematically studying the whole series of (Tb,Ho,Sc)V$_6$Sn$_6$ compounds, we also demonstrate its resilience against the onset of a charge ordered phase, a distinctive feature of many recently discovered kagome metals~\cite{Neupert_2022,arachchige2022,teng2022discovery}. Beside unveiling the interplay between many-body electronic states and topology in this class of materials, our work is testament of the first experimental measurements of a spin-Berry curvature in real quantum systems. Indeed, the detection of energy- and momentum-resolved finite Berry-curvature signal was hitherto limited to cold atoms experiments, which unveiled the deep relationship between topology and flatness in optical lattices \cite{Flaschner_2016}. Furthermore, in that context, flat bands serve as a platform for emergent correlated phases and their simplicity can advance the understanding of the physics that occurs in argon ice, Landau levels, and twisted Van der Waals bilayers \cite{Kruchkov_2022}. Here, we extend this context to real solid state systems for the first time.

TbV$_6$Sn$_6$, see Fig.~1{\bf a},  is a kagome system belonging to the “166” family of rare-earth kagome metals, along with GdV$_6$Sn$_6$ and HoV$_6$Sn$_6$~\cite{Pokharel_2021,Peng_2021, Ishikawa_2021, Rosenberg_2022}. It exhibits a uniaxial ferromagnetic transition at 4.1K with a substantial anisotropy in the magnetic susceptibility, suggesting a ferromagnetic alignment of Tb$^{3+}$ 4$f$ moments perpendicularly to the V kagome layers~\cite{Rosenberg_2022}. The DFT bulk electronic band structure is shown in Fig.~1{\bf b}. It is characterized by prominent features hinting toward a non-trivial topology. Dirac-like dispersions appear at the K points of the Brillouin zone (BZ) and contribute to the metallic character of the material. In addition, two flat bands are visible below and above the chemical potential, as highlighted by the yellow color proportional to the band- and momentum-resolved density of states $\rho_{n{\bf k}} \sim 1/v_{n{\bf k}}$, where $v_{n{\bf k}}$ represents the electronic velocity. Both the Dirac cones and the flat band around -1 eV are spectroscopically detectable by ARPES owing to their occupied character. As shown in Fig.~1{\bf c}, which zooms in the red rectangles (1) and (2) of Fig.~1{\bf b}, Dirac cones are gapped out by spin-orbit coupling. The SOC permits direct gaps between bands throughout the BZ, which in turn allow defining the $\mathbb{Z}_2$ topological invariant for the occupied bands using parity products at time-reversal invariant momenta~\cite{Fu_2007}. As for GdV$_6$Sn$_6$~\cite{Pokharel_2021,Hu_2022}, we find $\mathbb{Z}_2 = 1$ for the bands around the Fermi level. We also highlight that, in the absence of SOC, the Dirac cones carry a finite Chern number $\mathcal{C}$, and are source and sink of finite Berry curvature $\Omega({\bf k})$, indeed resulting in topologically protected arcs on the surface of this class of materials, as Fig.~1{\bf g} shows.

Upon cleavage, two terminations are possible, namely a Sn-terminated and a mixed V/Sn-terminated surface plane, with the latter being characterized by V atoms arranged into a kagome pattern. To determine the type of termination, we acquired the Sn core levels alongside the ARPES, finding a good agreement with previous works on the sister compound GdV$_6$Sn$_6$~\cite{Peng_2021}. Specifically, as shown in Fig.~1{\bf d}, the Sn termination exhibits the presence of two extra peaks at lower binding energies in the Sn $4d$ core levels, compatible with the corresponding Sn-derived surface components. The kagome termination, on the other hand, shows a significantly different line shape, featuring an asymmetric profile and a shift of about 0.43 eV towards higher binding energy values. These differences are attributed to the different local atomic environment present at the two surfaces~\cite{Peng_2021}. The measured Fermi surfaces (Fig.~1{\bf e}-{\bf f}) for both terminations agree well with the calculated ones (Supplementary Figure 2), and the typical kagome motif of corner-sharing triangles is also recognizable in reciprocal space.

The surface states at the kagome termination have minimal separation from the bulk continuum \cite{Hu_2022, Pokharel_2021}. This makes their spin-resolved measurements challenging, since the bulk contribution is intense and makes the spin-polarised signal too weak to be observed (see also Supplementary Figure 3 for additional spectra). However, the Sn-terminated surface features well-separated surface states, which allow the spin-ARPES experiments to be performed more easily, offering the perfect playground to investigate the topological properties of this system. We will focus on this termination when discussing the topological character of the surface states of TbV$_6$Sn$_6$. In contrast, the correlated flat band topology is not termination-dependent. Such a band, in fact, is a feature inherent the bulk kagome geometry, making this study of broad interest for kagome lattices in general and offering a comparative parallel case study for other systems. In addition, to determine the topology of the gap between the Dirac state and the flat band, the experimental measurement of the spin-Berry curvature is essential, being all the bands forming such a gap spin-degenerate, thus inaccessible by standard spin-ARPES.

The high-resolution electronic structure of the Sn termination is shown in Fig.~2. By using both linear vertical and horizontal light polarization ($E_s$ and $E_p$, respectively; see also the experimental setting in Fig.~2{\bf a}), we detect a plethora of interesting electrons' features, as shown in Fig.~2{\bf b}-{\bf c}: multiple Dirac dispersions, van Hove singularities, and the correlated flat band, which are a hallmark of kagome materials, are well identified. In particular, the van-Hove singularities are located slightly above the Fermi energy and at about -0.4 eV (labelled VH$_1$ and VH$_2$ in Fig.~2{\bf f}, respectively) and are visible in both theory and experiment. In addition, the Dirac cones form at the K point at -1.5 eV (D$_1$) and -0.3 eV (D$_2$) binding energy (see Fig.~2{\bf b},{\bf f}). The latter set of bands evolves symmetrically across the BZ with a quadratic minimum (QM in Fig.~2{\bf b}) at the zone center. According to a basic first-nearest-neighbors tight-binding calculation (see also Fig.~4{\bf a}), such a quadratic minimum is pinned to the flat band. We clearly detect both features in Fig.~2{\bf d} and relative energy distribution curve (blue and green curves) in Fig.~2{\bf e}. We also notice that the flat band is only visible with $E_s$ polarised light, whilst the quadratic minimum intensity disappears at the zone centre due to photoemission matrix elements. Nonetheless, the quadratic minimum is well identified in the energy distribution curve acquired with $E_p$ polarization (blue curve), for which, instead, the flat band is not visible. Thus, the combination of $E_s$ and $E_p$ allows us to visualize both the quadratic minimum and the flat band, and estimate the SOC-induced gap between them to be $\sim 60$ meV. Importantly, the presence of SOC mixes the orbital character across the opening point at $\Gamma$, and induces a finite spin-Berry curvature (see again Fig.~4{\bf a} and further discussion).

Close to the Fermi level, our ARPES measurements and DFT calculations permit to investigate the topological nature of the 166 family. We focus on the surface states originating from the multiple gapped Dirac cones described in Fig.~1{\bf b}-{\bf c}. In previous literature, theoretical works have attempted a topological classification for the $\mathbb{Z}_2$ invariant of the surface states across the BZ. According to Hu et al.~\cite{Hu_2022}, topological surface states in GdV$_6$Sn$_6$ are expected to cover a significant portion of the BZ, bridging a large bulk gap across $\Gamma$. In contrast to these data, our ARPES spectra in Fig.~2{\bf f} does not reveal any surface states through the Fermi level around the $\Gamma$ point, likely due to a different chemical potential in TbV$_6$Sn$_6$ which pushes them into the unoccupied region of the electronic density of states. This observation is consistent with our slab-calculation of the electronic structure in Fig.~2{\bf g}, which, differently from the spectrum in Fig.~1{\bf g}, fully accounts for the structural relaxation of atoms at the Sn-terminated surface, where these states are present. They are primarily unoccupied, but a portion of them bridges the gap below the Fermi energy along the $\bar{\Gamma}$-$\bar{\text{M}}$ line.
We also notice that in Fig.~2{\bf g}, these surface states form a small electron pocket close to the center of the BZ. This pocket is absent in ARPES. Nonetheless, the presence of these surface states close to the K point is enforced by topology, because they originate as Fermi arcs from the Dirac cones. Therefore, in this region, we will seek the signature of the spin-polarized feature in ARPES, because these states are well accessible there.

To experimentally verify this, we first calculated the expected spin-texture of the Sn surface states. The calculated spin-resolved electronic structure is shown in Fig.~3{\bf a}. We found that $S_y$ gives the most significant and only non-zero contribution along the $\bar{\Gamma}$-$\bar{\text{K}}$-$\bar{\text{M}}$ $k_x$ direction (See Supplementary Figure 9 for the $S_x$ and $S_z$ components). This result demonstrates the spin-momentum locking of the spin-texture and its non-trivial origin. Using the ARPES analyzer's deflectors, we measured the spin-ARPES signal along $\bar{\Gamma}$-$\bar{\text{K}}$-$\bar{\text{M}}$ at specific momenta for both positive and negative ${\bf k}$ values (colored vertical bars in Fig.~3{\bf a}). In this way, not only can the theoretical predictions be proved, but the time-reversal symmetry constraint can also be verified by keeping the same matrix elements. The spin-ARPES data for $S_y$ are shown in Fig.~3{\bf b}-{\bf e} (see Supplementary Information for details about data normalization). A clear spin-resolved signal in the proximity of the Fermi level confirms a non-vanishing spin-polarization typical of spin-polarized states. In addition, the spin sign reverses with the momentum ${\bf k}$, guaranteeing the time-reversal symmetry of the system. This important aspect is also compatible with the sample being well above the magnetic transition temperature (the measurements were indeed performed at 77 K). Additionally, the Tb $4f$-levels are well separated from the near-Fermi energy region, and do not hybridize significantly with the measured surface states (see also Supplementary Figure 4).

The energy distribution curves of Fig.~3{\bf b}-{\bf e} unambiguously demonstrate the spin-polarized character of the surface states in TbV$_6$Sn$_6$. In the Supplementary Figure 5, a thicker energy distribution curve map is also available for completeness. We notice that of the full set of states present in the DFT calculations, we are only able to resolve those which are well-separated from the bulk electronic structure, and that appear more prominent in intensity within the gap, still providing sufficient evidence for the spin-momentum locking expected for these spectrocopic features. As such, this makes our finding relevant within the framework of transport experiments in kagome lattices~\cite{Lima_2022, Chen_2021, Shuo_2020, Zeng_2022, Liu_2018, Neupert_2022, Owerre_2017, Heumen_2021}. 

Standard spin-ARPES, on the other hand, cannot be used to prove the topological character of the gap between the correlated kagome flat band and the quadratic minimum, because those bands are spin-degenerate. Theoretically, one can access the topological character of the gap by calculating the spin-Berry curvature of this system. The Chern number $\mathcal{C}$ of each band forming such a gap is identically zero owing to the combined action of inversion (which gives $\Omega_{n\uparrow}({\bf k}) = \Omega_{n\uparrow}(-{\bf k})$ for the Berry curvature) and time-reversal symmetry (which enforces $\Omega_{n\uparrow}({\bf k}) = \Omega_{n\downarrow}(-{\bf k})$). Nonetheless, one can expect a finite spin-Berry curvature $\Omega_z^{S_z}({\bf k})$ due to the action of SOC. At the level of a simple first-nearest-neighbors tight-binding model with hopping amplitude $t$, SOC opens a gap at the $\Gamma$ point between the parabolic dispersion from the Dirac band $(1)$ and the flat band $(2)$, as shown in Fig.~4{\bf a}.
The opening of a gap is in general associated with the appearance of a finite dispersion for the flat band~\cite{Tang_2011,NagaosaPRB2019}, and SOC can be thought of as a perturbation breaking the real-space topology that protects the band touching in generic frustrated hopping models~\cite{BalentsPRB2008}.
In modern language, kagome spectra without a mathematically flat band cannot be derived as one of the two isospectral partners of a supersymmetric bipartite graph with finite Witten index~\cite{TrebstSUSY}.
It results then that the electronic states around the opened gap feature a finite spin-Berry curvature, leading to a non-trivial spin-Chern number for the weakly-dispersing flat band itself.
Similar conclusions hold also for TbV$_6$Sn$_6$, as we show in Fig.~4{\bf b} (top panel), even though the band structure is much more complex than the simplified picture shown by the tight-binding model (notice, for instance, the presence of a spectator band carrying vanishing spin-Berry curvature at $\Gamma$). In addition, away from the flat band region around the $\Gamma$ point, our theoretical calculations reveal enhanced spin-Berry curvature at every SOC-induced avoided crossing between two spin-degenerate bands.

Sch\"{u}ler et al.~\cite{schuler_2020} recently proposed a methodology based on circular dichroism (CD) and spin-ARPES to directly address a signal proportional to the spin-Berry curvature $\Omega_z^{S_z}({\bf k})$ of quantum materials, thus allowing us to disentangle between trivial and non-trivial topology. CD has already been used in ARPES experiments to get information about the Berry curvature~\cite{Uenzelmann_2021,Cho_2021}. However, the presence of inversion symmetry and SOC in TbV$_6$Sn$_6$ requires separating between the two spin-channels, {\it i.e.} $\pm S_z$, as we show in Fig.~4{\bf c}-{\bf d}. In addition, to ensure that any geometrical contribution will not artificially alter our measurements, we measure the CD at the BZ center, where the geometrical contribution is exactly zero, and so it is the spin-integrated dichroic signal. Being at the BZ center also has the advantage that this point is time-reversal symmetric; thus, we are able to detect a signal reversal for the plus and minus components of the spin. Our measurements, shown in Fig.~4{\bf e}, demonstrate a substantial non-zero signal for each spin species of the quadratic minimum ($\sim -1$ eV) and for electronic bands at slightly higher binding energies where also a finite spin-Berry curvature appears as a result of SOC ($<\sim -1.2$ eV), with a reversal between spin up and down channels (red and blue, respectively). The spin-Berry curvature contribution at -1 eV comes from the quadratic minimum and this can be understood by looking at the spin-integrated CD ARPES (see Supplementary Figure 1) when compared to the measurements performed with $E_s$ and $E_p$ light polarization: the circular light has matrix elements similar to $E_p$, thus the CD ARPES results in a vanishing spectral weight for the flat band and a clear signal for quadratic minimum, at the BZ centre. Thus, we can attribute the strong intensity peaked at -1 eV of Fig.~4{\bf e} to the spin-Berry curvature of the quadratic minimum. Similar conclusions can we drawn also for HoV$_6$Sn$_6$, as we experimentally show in Fig.~4{\bf f}-{\bf g}. 
It is worth stressing that by changing photon energy, we were not able to resolve the flat band by using circularly polarized light; at the contrary, in this experimental configuration, the quadratic minimum was the only resolved feature. This result agrees with our calculations of Fig.~4{\bf b} (top panel) that suggest a finite spin-Berry curvature contribution around the SOC gap for the visible band $(1)$ forming the quadratic minimum. Our analysis is the first proof of a topological gap in a kagome metal. In addition, our CD measurements in Fig.~4{\bf e} unveil a large signal at energies below the flat band region. This result is again supported by our first-principles calculations of $\Omega_z^{S_z}({\bf k})$, as we show in Fig.~4{\bf b} (bottom panel), where the electronic states around $-1.3$ and $-1.4$ eV are characterized by an enhanced spin-Berry curvature.

Kagome metals are also getting much attention since they represent the perfect playground for several intertwined many-body orders~\cite{Neupert_2022}. Unconventional charge density wave (CDW) is one of these. Its origin, whether it arises from the electron-phonon coupling, electron-electron interaction, or a combination thereof, is still a matter of debate. Differently from TbV$_6$Sn$_6$ and HoV$_6$Sn$_6$, ScV$_6$Sn$_6$ shows a CDW phase below $T_{CDW}\sim 92$K, characterized by a distinctly different structural mode than that observed in the archetypal AV$_3$Sb$_5$ (A = K, Rb, Cs) compounds~\cite{arachchige2022} and FeGe~\cite{teng2022discovery}. In Fig.~4{\bf h}-{\bf i} we present our spin-resolved CD results for ScV$_6$Sn$_6$ at low-temperature, {\it i.e.} inside the charge ordered phase, whereas in the Supplementary Figure 10 we show that, besides an increase in the noise level due to thermal broadening, the aforementioned CD results are identical above and below $T_{CDW}$. Clearly, around -1 eV, we see a net spin asymmetry that reverses signs when the light polarization is changed. This result unambiguously demonstrates that the spin-Berry curvature is robust against the onset of the ordered phase, and that SOC-induced energy scale associated to the appearance of a finite spin-Berry curvature is larger than that correlated to the CDW symmetry breaking. Interestingly, our first-principles calculations, shown in Supplementary Figure 8, reveal that, upon unfolding the band structure of the distorted ScV$_6$Sn$_6$ onto the primitive unit-cell, the CDW distortion affects only marginally the electronic properties. This is in striking contrast to the effect of the CDW order in AV$_3$Sb$_5$ compounds, where sizable band gaps open around the chemical potential~\cite{Kang2022, Consiglio_2022}.


In conclusion, we have demonstrated the topological nature of XV$_6$Sn$_6$ kagome metals by exploiting the combination of spin-ARPES and DFT calculations and leveraging sensitivity at the multiple energy scales relevant to kagome systems. Beside unveiling a net spin-polarization of the surface states of TbV$_6$Sn$_6$, which originates from the non-trivial $\mathbb{Z}_2$ invariant of the SOC-induced bulk gaps close to the Fermi energy, we have crucially shown that the correlated flat band region is characterized by a finite spin-Berry curvature, establishing for the first time its topological character. In addition, we have reported the resilience of the non-trivial topology against the onset of the charge ordered phase in ScV$_6$Sn$_6$, revealing its ubiquitous nature across the series. This will motivate the investigation of the spin-Berry curvature in other kagome metals as well, such as AV$_3$Sb$_5$ and FeGe, where a non-trivial flat band separation has also been predicted and in which the charge order has a strong effect on the electronic properties. Within a more general perspective, our work constitutes the first evidence of the multidimensional topological nature, {\it i.e.} from surface to bulk states, of the “166” kagome family. It ultimately establishes these systems as a new domain for correlated topological metallicity with a nontrivial spin-Berry curvature of the wavefunctions manifold.\\

\noindent{\bf Acknowledgements}\\
The authors are grateful to Oleg Janson and Tilman Schwemmer for insightful discussions.
The research leading to these results has received funding from the European Union's Horizon 2020 research and innovation program under the Marie Sk{\l}odowska-Curie Grant Agreement No. 897276 (D.D.S.). We gratefully acknowledge the Gauss Centre for Supercomputing e.V. (https://www.gauss-centre.eu) for funding this project by providing computing time on the GCS Supercomputer SuperMUC-NG at Leibniz Supercomputing Centre (https://www.lrz.de). G.S., R.T., P.E., S.E. and A.C. are grateful for funding support from the Deutsche Forschungsgemeinschaft (DFG, German Research Foundation) under Germany's Excellence Strategy through the W\"urzburg-Dresden Cluster of Excellence on Complexity and Topology in Quantum Matter ct.qmat (EXC 2147, Project ID 390858490), through QUAST FOR 5249 (Project No. 449872909) as well as through the Collaborative Research Center SFB 1170 ToCoTronics (Project ID 258499086). This work has been performed in the framework of the Nanoscience Foundry and Fine Analysis (NFFA-MUR Italy Progetti Internazionali) facility (G.P. and G.R.). I.Z. acknowledges the support from US Department of Energy Early Career Award DE-SC0020130. The Flatiron Institute is a division of the Simons Foundation. S.D.W. and G. Po. acknowledge support via the UC Santa Barbara NSF Quantum Foundry funded via the Q-AMASE-i program under award DMR-1906325. C.B. and P.K. gratefully acknowledge support from The Leverhulme Trust through project RL-2016-006. F.M. greatly acknowledges the SoE action of pnrr, number SOE\_0000068.\\


\noindent{\bf Author contributions}\\
D.D.S, G.Pa., G.S, and F.M. conceived and designed the project. G.Po. and S.W. grew the crystals. F.M., C.B. and P.C. carried out the spin-ARPES and ARPES measurements. D.D.S., G.S., S.E., P.E. and A.C. performed the numerical calculations. All the authors participated in the discussion and contributed to the writing of the manuscript.\\

\noindent{\bf Competing financial interests}\\
The authors declare no competing financial interests.\\

\noindent{\bf Figure Legends/Captions}\\

\noindent{\bf Figure 1: Crystal structure, bulk electronic properties and surface terminations of TbV$_6$Sn$_6$ kagome metal.} {\bf a} Crystal structure of TbV$_6$Sn$_6$ showing top and side views of the unit cell. {\bf b} Bulk electronic structure along the $\Gamma$-K-M direction in the presence of spin-orbit coupling. The electronic states are colored by the band and momentum resolved density of states, with yellow highlighting a large contribution. {\bf c} Zoom-in of red boxes (1) and (2) in panel {\bf b}. Red and blue bands refer to calculations with and without SOC, respectively. {\bf d} Sn $4d$ core level spectroscopy for the kagome-terminated (green) and Sn-terminated (red) surface of TbV$_6$Sn$_6$. {\bf e}-{\bf f} ARPES  Fermi surfaces for the kagome and Sn termination of TbV$_6$Sn$_6$, respectively. {\bf g} Spectral function of the (001) surface Green's function for the Sn termination in the absence of SOC.\\

\noindent{\bf Figure 2: Spectroscopy of the surface states and of the flat band.} {\bf a} Experimental ARPES setting of linear vertical $(E_p)$ and horizontal $(E_s)$ light polarizations. $E_s$ is fully in-plane and parallel to the x-axis. $E_p$ has an incidence angle of $45^{o}$, thus with $50\%$ out-of-plane contribution and $50\%$ in-plane contribution parallel to the y-axis. {\bf b-c} ARPES spectra recorded with $E_p$ and $E_s$ lights, respectively. {\bf d} Enlarged view of the red box in panel {\bf c} highlighting the dispersion of the flat band around the $\Gamma$ point at $\sim -1$ eV of binding energy. {\bf e} energy distribution curve along the green line of panel {\bf d} collected with $E_s$ polarization (green) and the same but for $E_p$ (blue curve). The former shows the flat band position for $k_{x,y}=0$, the latter instead has prominent intensity in correspondence of the quadratic minimum, as also visible in panels {\bf b} and {\bf c}. {\bf f} Zoom-in of the ARPES data in the proximity of the Fermi level showing the most intense Sn-derived surface states (indicated by arrows labelled (1) and (2)) and also the van-Hove singularities VH$_1$ and VH$_2$ at the M point, with the maximum of VH$_1$ slightly above the Fermi level. {\bf g} First-principles electronic structure of a finite slab of TbV$_6$Sn$_6$ upon structural relaxation of the atoms at the Sn termination.\\

\noindent{\bf Figure 3: Spin-texture of topological surface states.} {\bf a} First-principles electronic structure of the Sn termination of a finite slab of TbV$_6$Sn$_6$ where the electronic states are colored by their $S_y$ character. {\bf b-e} $S_y$ Spin-resolved ARPES energy distribution curves at the specific momenta highlighted in panel {\bf a} by the colored vertical bars. The grey arrows refer to the energy regions where a finite spin asymmetry for the surface states is measured.\\

\noindent{\bf Figure 4: Topology of the flat band's region.} {\bf a} Band structure of a first-nearest-neighbor tight-binding model on the kagome lattice with SOC. The color highlights the finite contribution from the spin-Berry curvature $\Omega_z^{S_z}({\bf k})$ around the SOC-induced gaps formed by bands $(1)$ and $(2)$. {\bf b} Same as {\bf a} but for a realistic first-principles tight-binding model of TbV$_6$Sn$_6$ around the flat band region (top panel) and at slightly higher binding energies (bottom panel). {\bf c} Circular right and {\bf d} circular left spin-ARPES energy distribution curves, collected for the spin--$\uparrow$ ($S_z>0$, red) and spin--$\downarrow$ ($S_z<0$, blue) channels. The energy distribution curves have been collected at the centre of the BZ with the light entirely in the sample's mirror plane. In these conditions, the geometrical contribution coming from the CD can be safely excluded. {\bf e} Extracted spin-polarization for the CD, showing a strong and finite contribution for the quadratic minimum around -1 eV as high as $\sim \pm$90$\%$, as well as at higher binding energies. In grey, we also reported the percentage contribution of the CD (CR-CL) for spin-integrated ARPES, showing fluctuations of maximum $\pm8\%$ in the region of interest. {\bf f}-{\bf g} and {\bf h}-{\bf i} Same as {\bf c}-{\bf d} for HoV$_6$Sn$_6$ and ScV$_6$Sn$_6$, respectively.\\


\vspace{1cm}
\noindent{\bf Methods}\\

\noindent{\bf Experimental details --}
Single crystals of XV$_6$Sn$_6$ (Tb, Sc, Ho) were grown using a flux-based growth technique as reported in the reference \cite{Pokharel_2022}. X (chunk, 99.9$\%$), V (pieces, 99.7$\%$), and Sn (shot, 99.99$\%$) were loaded inside an alumina crucible with the molar ratio of 1:6:20 and then heated at 1125$^{o}$C for 12 h. Then, the mixture was slowly cooled to 780$^{o}$C at a rate of 2$^{o}$C/h. Thin plate-like single crystals were separated from the excess Sn-flux via centrifuging at 780$^{o}$C. The samples were cleaved in ultrahigh vacuum (UHV) at the pressure of $1 \times 10^{-10}$ mbar. The spin-ARPES data were acquired at the APE-LE end station (Trieste) using a VLEED-DA30 hemispherical analyzer. The energy and momentum resolutions were better than 12 meV and 0.02 \AA$^{-1}$, respectively. The temperature of the measurements was kept constant throughout the data acquisitions (16 K and 77 K), above the magnetic transition of the system ($< 5$ K). Both linear and circular polarized lights were used to collect the data from the APE undulator of the synchrotron radiation source ELETTRA (Trieste).

\noindent{\bf Theoretical details --}
We employed first-principles calculations based on the density functional theory as implemented in the Vienna ab-initio simulation package (VASP)~\cite{Kresse_1996}, within the projector-augmented plane-wave (PAW) method~\cite{PAW}. The generalized gradient approximation as parametrized by the PBE-GGA functional for the exchange-correlation potential is used~\cite{pbe} by expanding the Kohn–Sham wave functions into plane waves up to an energy cutoff of 400 eV. We sample the Brillouin zone on an $12 \times 12 \times 6$ regular mesh by including SOC self-consistently. For the calculation of the surface spectral function, the Kohn–Sham wave functions were projected onto a Tb $d$, V $d$ and Sn $s$, $p$-type basis.
The calculation of the spin-Berry curvature requires a Wannier Hamiltonian where the lattice symmetries are properly enforced. For this reason, we used the full-potential local-orbital (FPLO) code~\cite{FPLO}, version 21.00-61 [https://www.FPLO.de/]. The spin-Berry curvature for band $n$ is then defined as

\begin{equation}
    \label{eqSBC}
    \Omega_{xy}^{z}({\bf k}) = \sum_{E_{n}>E_{m \ne n}} \frac{ \langle n | v_{s,x}^{z}| m \rangle \langle m | v_y | n \rangle (x \leftrightarrow y)}{( E_{n{\bf k}} - E_{m{\bf k}} )^2} \, ,
\end{equation}

\noindent with the spin operator $\sigma_z$ and velocity operator $v_i = \frac{1}{\hbar} \partial H / \partial k_i$ ($i=x,y$). $| n {\bf k} \rangle$ is the eigenvector of the Hamiltonian $H$ with the eigenvalue $E_{n{\bf k}}$. Eq.~\ref{eqSBC} is computed by using our in-house post-wan library (see Code availability statement).\\

\noindent{\bf Data availability} Source data are available at the DOI:10.5281/zenodo.7787937\\

\noindent{\bf Code availability} Our in-house post-wan library used to compute Berry curvature-related quantities can be downloaded from https://github.com/philipp-eck/post\_wan.\\

\newpage

\begin{figure*}[!t]
\centering
\includegraphics[width=\textwidth,angle=0,clip=true]{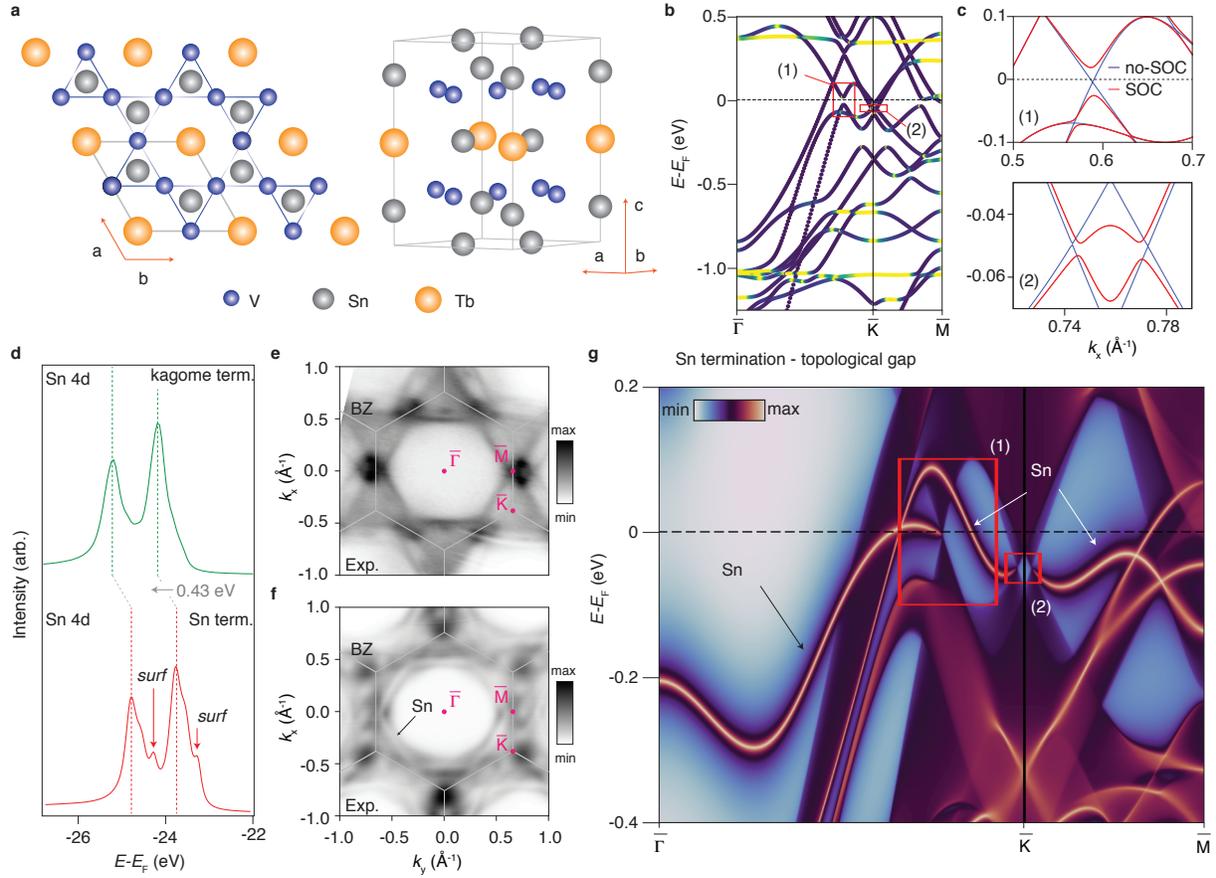}
\caption{{\bf Crystal structure, bulk electronic properties and surface terminations of TbV$_6$Sn$_6$ kagome metal.} {\bf a} Crystal structure of TbV$_6$Sn$_6$ showing top and side views of the unit cell. {\bf b} Bulk electronic structure along the $\Gamma$-K-M direction in the presence of spin-orbit coupling. The electronic states are colored by the band and momentum resolved density of states, with yellow highlighting a large contribution. {\bf c} Zoom-in of red boxes (1) and (2) in panel {\bf b}. Red and blue bands refer to calculations with and without SOC, respectively. {\bf d} Sn $4d$ core level spectroscopy for the kagome-terminated (green) and Sn-terminated (red) surface of TbV$_6$Sn$_6$. {\bf e}-{\bf f} ARPES  Fermi surfaces for the kagome and Sn termination of TbV$_6$Sn$_6$, respectively. {\bf g} Spectral function of the (001) surface Green's function for the Sn termination in the absence of SOC.}
\label{fig1}
\end{figure*}

\newpage

\begin{figure*}[!ht]
\centering
\includegraphics[width=\textwidth,angle=0,clip=true]{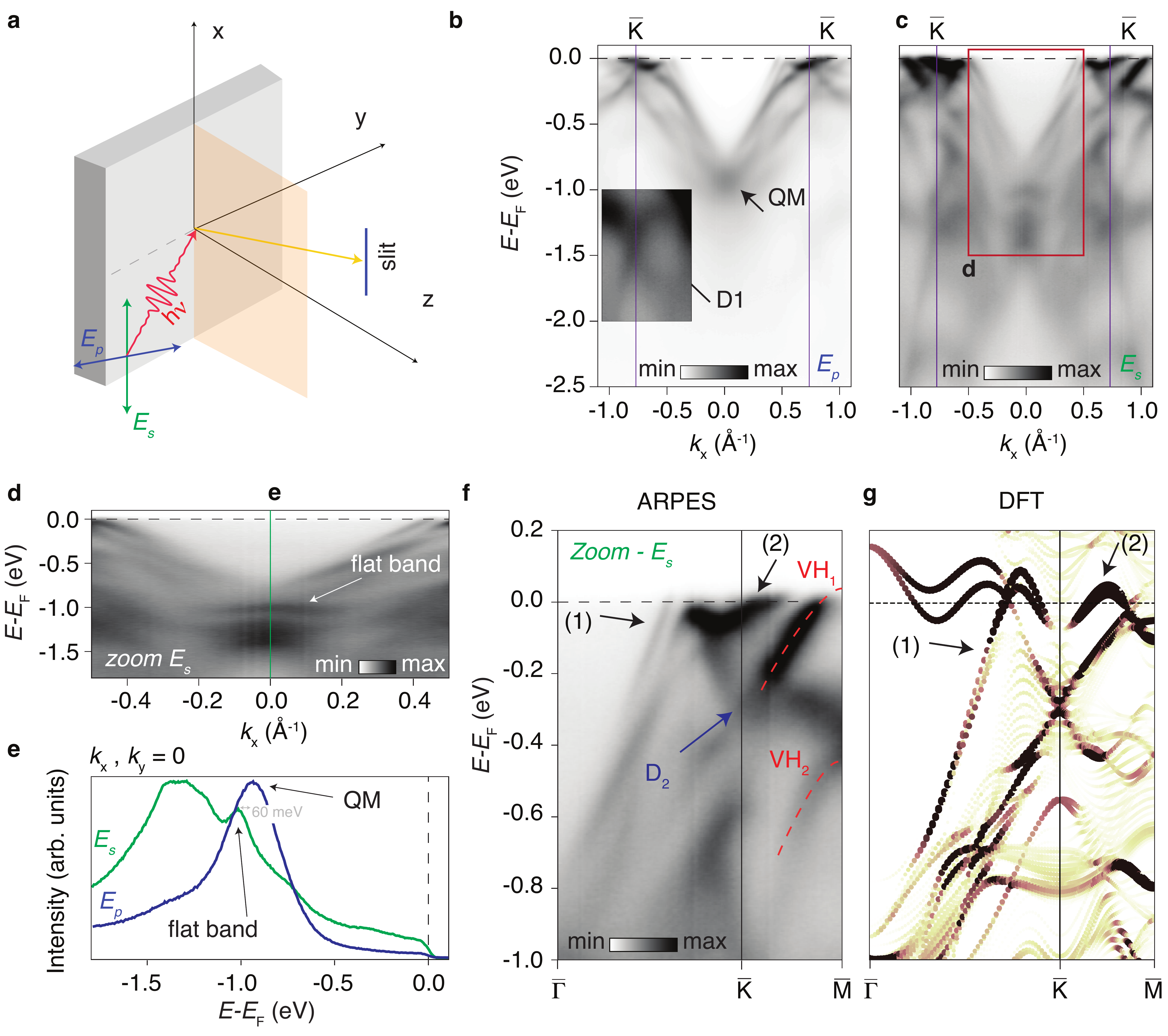}
\caption{{\bf Spectroscopy of the surface states and of the flat band.} {\bf a} Experimental ARPES setting of linear vertical $(E_p)$ and horizontal $(E_s)$ light polarizations. $E_s$ is fully in-plane and parallel to the x-axis. $E_p$ has an incidence angle of $45^{o}$, thus with $50\%$ out-of-plane contribution and $50\%$ in-plane contribution parallel to the y-axis. {\bf b-c} ARPES spectra recorded with $E_p$ and $E_s$ lights, respectively. {\bf d} Enlarged view of the red box in panel {\bf c} highlighting the dispersion of the flat band around the $\Gamma$ point at $\sim -1$ eV of binding energy. {\bf e} EDC along the green line of panel {\bf d} collected with $E_s$ polarization (green) and the same but for $E_p$ (blue curve). The former shows the flat band position for $k_{x,y}=0$, the latter instead has prominent intensity in correspondence of the QM, as also visible in panels {\bf b} and {\bf c}. {\bf f} Zoom-in of the ARPES data in the proximity of the Fermi level showing the most intense Sn-derived surface states (indicated by arrows labelled (1) and (2)) and also the van-Hove singularities VH$_1$ and VH$_2$ at the M point, with the maximum of VH$_1$ slightly above the Fermi level. {\bf g} First-principles electronic structure of a finite slab of TbV$_6$Sn$_6$ upon structural relaxation of the atoms at the Sn termination.}
\label{fig2}
\end{figure*}

\newpage

\begin{figure*}[!ht]
\centering
\includegraphics[width=\textwidth,angle=0,clip=true]{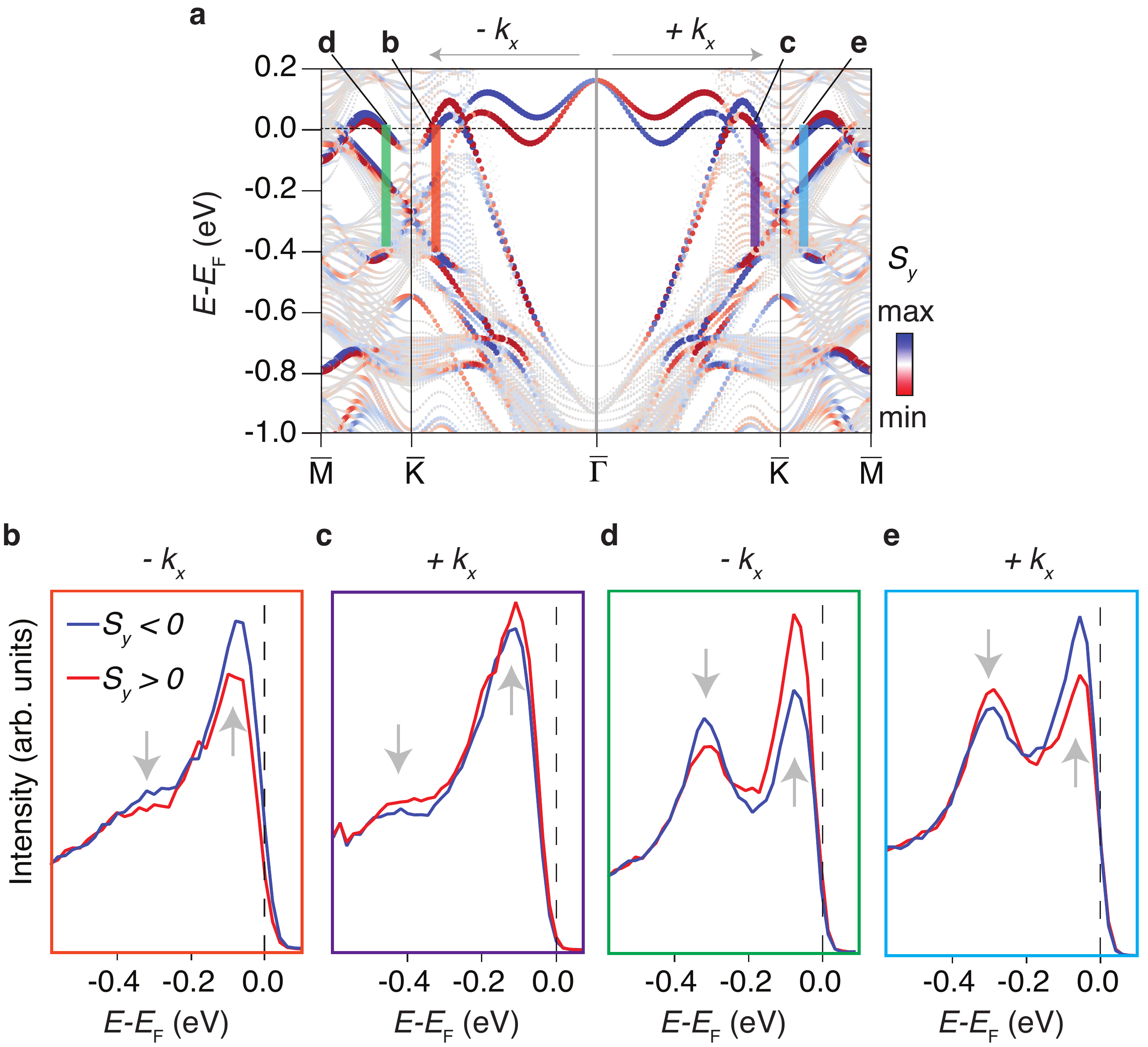}
\caption{{\bf Spin-texture of topological surface states.} {\bf a} First-principles electronic structure of the Sn termination of a finite slab of TbV$_6$Sn$_6$ where the electronic states are colored by their $S_y$ character. {\bf b-e} $S_y$ Spin-resolved ARPES EDCs at the specific momenta highlighted in panel {\bf a} by the colored vertical bars. The grey arrows refer to the energy regions where a finite spin asymmetry for the surface states is measured.}
\label{fig3}
\end{figure*}

\newpage

\begin{figure*}[!t]
\centering
\includegraphics[width=0.95\textwidth,angle=0,clip=true]{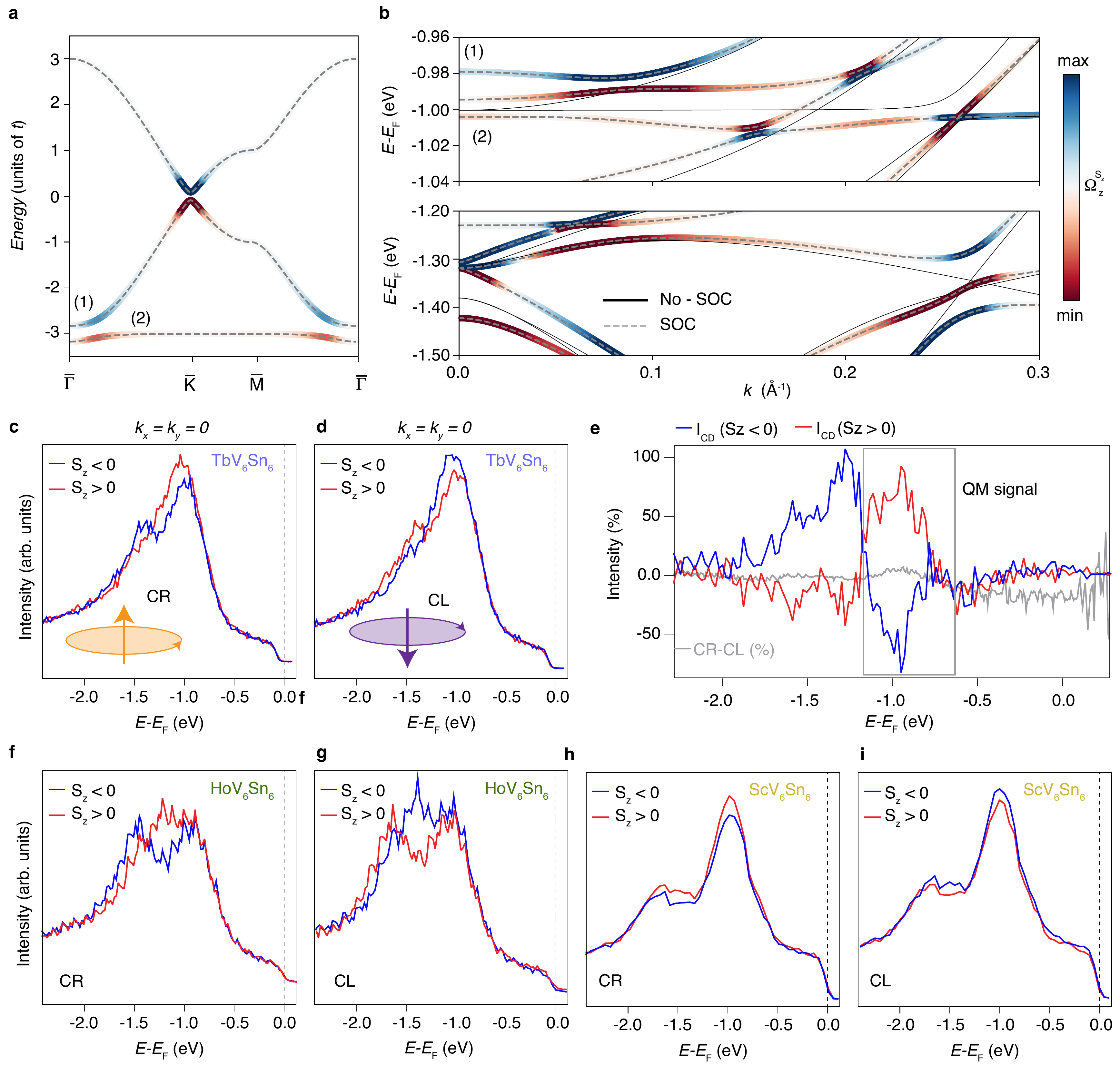}
\caption{{\bf Topology of the flat band's region.} {\bf a} Band structure of a first-nearest-neighbor tight-binding model on the kagome lattice with SOC. The color highlights the finite contribution from the SBC $\Omega_z^{S_z}({\bf k})$ around the SOC-induced gaps formed by bands $(1)$ and $(2)$. {\bf b} Same as {\bf a} but for a realistic first-principles tight-binding model of TbV$_6$Sn$_6$ around the flat band region (top panel) and at slightly higher binding energies (bottom panel). {\bf c} Circular right and {\bf d} circular left spin-ARPES EDCs, collected for the spin--$\uparrow$ ($S_z>0$, red) and spin--$\downarrow$ ($S_z<0$, blue) channels. The EDCs have been collected at the centre of the BZ with the light entirely in the sample's mirror plane. In these conditions, the geometrical contribution coming from the CD can be safely excluded. {\bf e} Extracted spin-polarization for the CD, showing a strong and finite contribution for the QM around -1 eV as high as $\sim \pm$90$\%$, as well as at higher binding energies. In grey, we also reported the percentage contribution of the CD (CR-CL) for spin-integrated ARPES, showing fluctuations of maximum $\pm8\%$ in the region of interest. {\bf f}-{\bf g} and {\bf h}-{\bf i} Same as {\bf c}-{\bf d} for HoV$_6$Sn$_6$ and ScV$_6$Sn$_6$, respectively.}
\label{fig4}
\end{figure*}

\end{document}